\documentclass[a4paper,10pt,twocolumn]{amsart}
\usepackage{fullpage}
\usepackage{amssymb,epsfig,amsmath,amsthm}
\usepackage[font=small,format=plain,labelfont=bf,up,textfont=it,up]{caption}
\usepackage{graphicx,subfig}

\graphicspath{{./Figs/}}
\title{A Bellman View of Jesse Livermore}
\author{Nick Polson\vspace{0.1cm}\\
{\tiny and}\vspace{0.1cm}\\
Jan Hendrik Witte}

\thanks{Nick Polson, Booth School of Business, University of Chicago, {\tt ngp@chicagobooth.edu}.\\
Jan Hendrik Witte, Mathematical Institute, University of Oxford, {\tt witte@maths.ox.ac.uk}}

\begin{document}

\maketitle

\newtheorem{theorem}{Theorem}[section]
\newtheorem{la}[theorem]{Lemma}
\newtheorem{cor}[theorem]{Corollary}
\newtheorem{remark}[theorem]{Remark}
\newtheorem{prob}[theorem]{Problem}	
\newtheorem{definition}[theorem]{Definition}
\newtheorem{alg}[theorem]{Algorithm}
\newtheorem{prop}[theorem]{Proposition}
\newtheorem{assumption}[theorem]{Assumption}
\newtheorem{example}[theorem]{Example}

\section{Introduction}

Richard Bellman's \emph{Principle of Optimality}, formulated in 1957, is the heart of dynamic programming, the mathematical
discipline which studies the optimal solution of multi-period decision problems.\bigskip

In his 1923 book \emph{Reminiscences of a Stock Operator}, the legendary trader Jesse Livermore gave a detailed account of his trading methods.\bigskip

In this article, we study some of Livermore's trading rules, and we show that many of them directly reflect in 
Bellman's \emph{Principle of Optimality}. Thus, in their strive for optimality, two of the greatest minds of the 20th century can be found to be neatly aligned.\bigskip

Richard Bellman's 1957 book on \emph{Dynamic Programming} introduces his
conceptual framework for the solution of multi-stage decision processes. While having multiple different mathematical formulations, the
problems studied by Bellman all share the following main characteristics.\bigskip

\begin{itemize}
 \item There is a system, characterised at each stage by a set of parameters and state variables. 
 \item At each stage of either process, we have a choice of a number of decisions.
 \item The effect of a decision is a transformation of the state variables.
 \item Past history is of no importance in determining future actions.
 \item The purpose is to maximise a function of the state variables. 
\end{itemize}
\bigskip

In Bellman's words, a ``\emph{Policy}'' is any rule for making decisions which yields an allowable sequence of decisions; and an ``\emph{Optimal Policy}'' is
a policy which maximises a pre-assigned function of the final state variables. For every problem with the above listed properties, Bellman establishes
the following rule.\bigskip

\emph{Bellman's Principle of Optimality:}\bigskip 

\begin{changemargin}{0.5cm}{0.5cm} 
\setlength{\parindent}{0pt}
{\footnotesize

``An optimal policy has the property that, whatever the initial state and initial decision, the remaining decisions must constitute an optimal policy with regard to the state resulting from the first decision.''

}
\end{changemargin}\medskip

It becomes immediately clear that Bellman's criteria for optimality are reminiscent of a trading process, 
and that, therefore, the \emph{Principle of Optimality} should also be applicable to trading.\bigskip

The \emph{Principle of Optimality} suggests that we study the $Q$-value matrix describing the 
value of performing action $a$ in our current state $s$, and then acting optimally henceforth. In this framework, let $Q(s,a)$ denote the set of values available from current state $s$ through action $a$; e.g., interpret $s$ as the agent's current wealth, and $a$ as a parametrisation of a long or short position (or any other action) he initiates.\bigskip

The current \emph{Optimal Policy} and \emph{Value Function} are given by
\begin{align*}
a^\star(s)  := &\ {\rm arg \; max}_a \; Q(s,a)\\
 {\rm and} \; \; V(s) :=&\ \max\nolimits_a Q(s,a) = Q(s,a^\star(s)),
\end{align*}
respectively. Let $s^\prime $ denote the next state of the system, and let $ r( a ,s , s^\prime ) $
be the reward of the next state $s^\prime$ given current state $s$ and action $a$.\bigskip

We would like to put the just introduced definitions into a sequential context. We will consider the path of a single market where trading takes place at discrete times. Time $t=0$ corresponds to the current time. The realised current state will be denoted by $s_t$, the current action by $a_t$, a future unobserved state by $S_{t+1}$, and optimal policy and value functions are defined by
\begin{align*}
a^\star(s_t)  =&\ {\rm arg \; max}_{a} \; Q_{t+1}(s_t,a)\\
 {\rm and} \; \; V(s_t) =&\ Q_{t+1}(s_t,a^\star(s_t)),
\end{align*}
respectively.\bigskip

The \emph{Principle of Optimality} now provides the key sequential identity of \emph{Dynamic Programming}, namely that
\begin{align}
&Q_{t+1} (s_t,a)\label{Q_OptPrin}\\
 =&\ \mathbb{E} \left[ r( a , s_t , S_{t+1} ) + V( S_{t+1} ) \mid s_t ,\, a \right].\nonumber
\end{align}
In Bellman's words: whatever today's state $s_t$, and whatever today's decision $a$, today's value $Q_{t+1} (s_t,a)$ is based on (expected) optimal decision making with regard to
the next state $S_{t+1}$ which results from today's state $s_t$ and decision $a$.\bigskip 

To find today's optimal action, one has to solve the equilibrium condition \eqref{Q_OptPrin} for the $Q$-matrix, and then read off the optimal action $a^\star(s_t)$ that attains 
$ {\rm arg \; max}_a \; Q(s_t,a) $. For simplicity, we assume that $ a_t \in \mathcal{A} $ takes only a finite set of possible values.\bigskip

(Equations \eqref{Q_OptPrin} and \eqref{Q_OptPrin_hat} also allow for inclusion of a discounting factor, which, if required, should be incorporated in $r( a , s_t , S_{t+1})$.)\bigskip

However, in reality, there's a caveat: to evaluate \eqref{Q_OptPrin} in his decision making, as the real world probabilities are unknown to him, the trader has to take expectations under his subjective probability distribution $q(S_t | s_t ,a )$, which describes his beliefs about the future path of state variables depending on his current wealth $s_t$ and his action $a$.\bigskip

Thus, instead of \eqref{Q_OptPrin}, the trader will attempt to solve
\begin{align}
&Q^q_{t+1} (s_t,a)\label{Q_OptPrin_hat}\\
=&\ \mathbb{E}^q\left[ r( a , s_t , S_{t+1} ) + V^q( S_{t+1} ) \mid s_t,\, a\right],\nonumber
\end{align}
where $\mathbb{E}^q[\cdot]$ and $ V^q(\cdot)$ denote probabilities taken with respect to the distribution $q(S_{t+1}| s_t,a)$ of the trader's beliefs.\bigskip

Within the Bellman and Livermore optimal framework, we note a number of compelling features, which we summarise in the following remarks.

\begin{remark}\label{remark1}
It is surprising how little effect the distinction between \eqref{Q_OptPrin} and \eqref{Q_OptPrin_hat} has on the actual trading process.
\end{remark}

\begin{remark}\label{remark1a}
Rules based on deviations between realised market prices and a trader's expectations have little place in assessing the optimal action $a^\star$: arguments such as ``\emph{sell} because prices went higher than \emph{my} expectations'' do not enter the picture, which is a version of Livermore's (1923) 
maxim that ``the market is never wrong''. Put simply, we should only worry about the optimal Bellman path of actions, or how we got there,
rather to act optimally from here on out.
\end{remark}

\begin{remark}\label{remark2}
A large part of the Bellman and Livermore optimal policy insight is that the traders subjective beliefs \eqref{Q_OptPrin_hat} must be updated \emph{conditionally} on observed market prices. Because the market has a superior information set when prices rise, 
the optimal action is to do nothing -- in Livermore's (1923) words, ``one should hope, not fear'' -- and when prices fall, one should think of selling -- ``one should fear, not hope''.
\end{remark}

We will look at the importance of Remarks \ref{remark1}, \ref{remark1a}, and \ref{remark2} in more detail in the next section.

\section{Trading Principles}

We discuss two of Jesse Livermore's main trading rules, hoping to provide a modest insight into his trading principles.

\subsection{Profits take care of themselves, losses never do}\label{LossesNeverDo_SubSec}

Suppose there are only two possible market positions, \emph{Long} and \emph{Neutral}, denoted by $a_L$ and $a_N$, respectively. \emph{Short} will be the reflection
of \emph{Long}.\bigskip

Suppose a trader initiates a long position $a_L$ at time $t$ because, given his current wealth $s_t$, he observes
\begin{align}
a_L
=&\ a^*(s_t)\label{LossesNeverDo_SubSec_Eq1}\\
=& {\rm arg \; max}_a\ Q^q_{t+1} (s_t,a)\nonumber\\
=&\ {\rm arg \; max}_a\ \mathbb{E}^q \big[ r( a , s_t , S_{t+1} ) + \nonumber\\
&\quad\quad\quad\quad\quad\quad\quad\quad V^q( S_{t+1} ) \mid s_t,\, a \big].\nonumber
\end{align}
But, suppose that, at time $t+1$, he finds that $r( a , s_t , s_{t+1} )<0$, and that his new wealth now is $S_{t+1}<s_t$. Then the trader has to evaluate whether to close his position (i.e., action $a_N$), or whether to hold on to his position (i.e., action $a_L$).\bigskip

If we assume that the decision whether to be in or out of the market is independent of the current wealth level $S_{t+1}$, then
\begin{equation*}
{\rm arg \; max}_a\ Q^q_{t+2}(S_{t+1},a) = {\rm arg \; max}_a\ Q^q_{t+2}(s_t,a).
\end{equation*}

We observe that, if acting under an unchanged subjective probability
\begin{equation*}
q=q(S_{t+1}| s_t, a), 
\end{equation*}
then he trader will proceed with $a_L$\,, since
$$
Q^{q}_{t+2}(S_{t+1},a_L) > Q^{q}_{t+2}(S_{t+1},a_N)
$$
as before.\bigskip

However, if we define
\begin{align*}
Q^{q,x_{t+1}}_{t+2} (S_{t+1},a)
:=&\ \mathbb{E}^{q} \big[ r( a , s_t , S_{t+2} ) +\\
&\quad\quad\quad V^{q}( S_{t+2} ) \mid s_t,\, a,\, x_{t+1} \big]
\end{align*}
for $a\in\{a_L,a_N\}$,
then the trader now supplements his rational (i.e., his subjective probability $q$) by the information contained in the most recent price move $x_{t+1}$.\bigskip

If we denote the trader's updated views by $q\cup\{x_{t+1}\}$, then, by a symmetry argument, we get that 
$$
Q^{q\cup\{x_{t+1}\}}_{t+2}(S_{t+1},a_L) < Q^{q\cup\{x_{t+1}\}}_{t+2}(S_{t+1},a_N),
$$
where, by assumption, $$r( a_L , s_t , S_{t+1} ) < 0 = r( a_N , s_t , S_{t+1} ).$$

We observe that, in absence of other external information besides the price move $x_{t+1}$, and being consistent with his own former rationale, the trader should 
take a \emph{Neutral} action and exit his position.\bigskip

In reality, due to cost of trading, the trader's reaction will not be immediate. However, the implication
of the principle of optimality is that the only thing of concern with is \emph{Selling} the losing position.\bigskip

The reflection of the above argument, in the 
case where $  r( a_L , s_t , S_{t+1} ) > 0$, shows that next period's optimal action is $a_L$ the same as the current period: the winners take care of themselves.\bigskip

\emph{In Livermore's (1923) words:}\bigskip 

\begin{changemargin}{0.5cm}{0.5cm} 
\setlength{\parindent}{0pt}
{\footnotesize

``Profits always take care of themselves, but losses never do. The speculator has to insure himself against considerable losses by taking the first small loss. In doing so, he keeps his account in order, so that, at some future time, when he has a constructive idea, he will be in a position to go into another deal, taking the same amount of stock he had when he was wrong.'' \medskip 

}
\end{changemargin}\medskip

We shall now look at a brief example, putting the just introduced concepts into practice.

\begin{changemargin}{0.5cm}{0.5cm} 
\setlength{\parindent}{0pt}
{\footnotesize

\begin{example}\label{CharlieStockMarket}
\upshape{
Consider a trader Jan, who is investing in Google (GOOG) shares. Suppose Jan's only counter-party is a broker called Theobald-Fritz, but whom we will nickname \emph{Hermes}\footnote{In classic Greek Mythology, \emph{Hermes} is the patron of travellers, herdsmen, poets, athletes, invention, \emph{trade}, and \emph{thieves}. The role played by Hermes here also strongly resembles Benjamin Graham's 1949 creation of \emph{Mr Market}, which is intentional.}.\bigskip

Suppose Jan has just purchased GOOG shares worth $\$1000$ from Hermes, i.e., we have $s_t=\$1000$. Suppose that Jan trades with Hermes daily, and that, every time he trades, he invests exactly $\$1000$ (independently of his net wealth), taking his profit/loss for the trade on the following day. Suppose further that the daily price movements of the GOOG shares are exactly $\pm 1\%$, where $p$ and $1-p$ are the real-world probabilities for an up or down move, respectively.\bigskip

Define $u:=\$10$ and $d:=-\$10$, and denote the decision to take a long (short) position by $a_L$ (by $a_S$). As $p$ is unknown to him, Jan has to make his trading decision based on his personal beliefs $1>q,1-q>0$. According to Jan's own estimate, $q>0.5$, and
\begin{align*}
& Q^q_{t+1} (s_t,a_L)\\
=&\ \mathbb{E}^q\left[ r( a , s_t , S_{t+1} ) + V^q( S_{t+1} ) \mid s_t,\, a=a_L\right]\\
>&\ \mathbb{E}^q\left[ V^q( S_{t+1} ) \mid s_t,\, a=a_N\right]\\
>&\ \mathbb{E}^q\left[ r( a, s_t , S_{t+1} ) + V^q( S_{t+1} ) \mid s_t,\, a=a_S\right],
\end{align*}
and, therefore, Jan his very happy with his newly purchased share of GOOG equity.\bigskip

Suppose in reality $p<0.5$, and that, the following day, Jan checks with Hermes, just to find that $S_{t+1}=s_t-d=\$990$ and $r( a_L , s_t , s_t-d)=-\$10<0$.\bigskip

Naturally, Jan is somewhat unimpressed with his results, and he initially blames Hermes. 
But, Jan has to make a decision: should he increase his holding GOOG equity following $a_L$?\bigskip

Jan's personal estimate $q>0.5$ is unchanged, so he is tempted to stay long and buy more. However, having carefully read Bellman and Livermore's books, 
Jan is also blissfully aware that this would not be in accordance with the optimal Bellman policy -- which
would have returned $r( a_S , s_t , s_t+d)=\$10>0$ up to this point. Therefore, realising that increasing his position would inadvertedly deviated from the optimal path, 
Jan decides to liquidate his position.} \bigskip

Of course, had Jan been profitable with $a_L$, the reverse argument would have held, and he would have kept his GOOG equity, enjoying the ride on the optimal 
Bellman trajectory (and avoiding any disputes with Hermes).

\end{example}
}
\end{changemargin}\medskip

\subsection{Don't average down}\label{Sec_DontAvDown} Averaging down is the practice of increasing one's position after taking a loss, in the hope of reaping the expected profit and recovering all previous losses.\bigskip

Strictly speaking, averaging down is already prohibited if losing positions are exited, which we covered in Section \ref{LossesNeverDo_SubSec}; however, the strategy is so popular that it warrants separate consideration.\bigskip

Using the notation of Section \ref{LossesNeverDo_SubSec}, suppose again that, based on \eqref{LossesNeverDo_SubSec_Eq1}, our trader holds a long position $a_L$ at time $t$, and that, at time $t+1$, he finds that $r( a_L , s_t , S_{t+1} )<0$, and that his new wealth now is $S_{t+1}<s_t$. If our trader thinks that an increased long position is in order, then, clearly, he must think that
\begin{align}
Q^{q'}_{t+2}(S_{t+1},a_L) > Q^{q'}_{t+2}(S_{t+1},a_N),\label{Sec_DontAvDown_Eq1}
\end{align}
where $q'$ denotes his updated personal probabilities.
But, in absence of other external information besides the price move $x_{t+1}$, we have $q'=q\cup\{x_{t+1}\}$, and, as already seen in Section \ref{LossesNeverDo_SubSec}, \eqref{LossesNeverDo_SubSec_Eq1} then implies
\begin{align}
Q^{q'}_{t+2}(S_{t+1},a_L) < Q^{q'}_{t+2}(S_{t+1},a_N),\label{Sec_DontAvDown_Eq2}
\end{align}
which means that \eqref{Sec_DontAvDown_Eq1} cannot be true.\bigskip

The contradiction between \eqref{Sec_DontAvDown_Eq1} and \eqref{Sec_DontAvDown_Eq2} is an interesting one, and slightly exceeds the implications of Section \ref{LossesNeverDo_SubSec}.
As acting based on $q'=q_t\cup\{x_{t+1}\}$ leads to \eqref{Sec_DontAvDown_Eq2}, we see that averaging down can only ever be justified if the trader believes to have obtained a new set of external information, exceeding what was learned from the latest price move $x_{t+1}$, and dominating the price  -- a very rare case indeed: generally, doubling up on a losing position is irrational.\bigskip

\emph{In Livermore's (1923) words:}\bigskip 
 
\begin{changemargin}{0.5cm}{0.5cm} 
\setlength{\parindent}{0pt}
{\footnotesize
 
``One other point: it is foolhardy to make a second trade, if your first trade shows you a loss. Never average losses. Let that thought be written indelibly upon your mind.''\bigskip 
 
``I have warned against averaging losses. That is a most common practice.
Great numbers of people will buy a stock, let us say at \$50, and two or three
days later if they can buy at \$47 they are seized with the urge to average down. [...]
If one is to apply such an unsound principle, he should keep on averaging
by buying 200 shares at \$44, then 400 at \$41, 800 at \$38, 1600 at \$35,
3200 at \$32, 6400 at \$29 and so on. How many speculators could stand such pressure?
Yet if the policy is sound it should not be abandoned. Of course, abnormal moves such as the
one indicated do not happen often. But it is just such abnormal moves which the speculator must guard against to avoid disaster.''\bigskip 
 
``So, at the risk of repetition, let me urge you to avoid averaging down. [...] Why send good money after bad? Keep that good money for another day. Risk it on something more attractive than an obviously losing deal.''\medskip
 
}
\end{changemargin}\medskip

\begin{remark}
An immediate corollary to Section \ref{Sec_DontAvDown} is that trading on the belief of a `true' value is dangerous. Almost always, the true value is established once, and then convergence is waited for. Adverse movements are interpreted as providing `better entry points', and are believed to `strengthen' the opportunity -- clearly, any such reasoning compounds the conflict between \eqref{Sec_DontAvDown_Eq1} and \eqref{Sec_DontAvDown_Eq2} severalfold, and should be strictly avoided.
\end{remark}

\begin{remark}
It is helpful to add that Jesse Livermore's original writings were independent of any specific market structure, but were presented to hold in generality, for \emph{any market}. Similarly, Richard Bellman's \emph{Principle of Optimality} applies to any multi-period decision making process. Therefore, the Bellman and Livermore optimal policy insight presented in this article applies to any financial transaction which takes place within an exogenously given market place of \emph{any form}.
\end{remark}

\section{Conclusion}

There is a saying, attributed to John Kenneth Galbraith, that \emph{``faced with the choice between changing one's mind and proving that there is no need to do so, almost everyone gets busy on the proof.''} In this article, we show that, within the Bellman and Livermore optimal policy insight, every view held should be updated conditionally on the latest available information; an unwillingness to \emph{learn}, as alluded to by Galbraith, is expected to be detrimental.\bigskip

On his website, Joe Fahmy gives a summary of Jesse Livermore's principles in his own words. His list, once again, nicely reflects the parallels between the Bellman and Livermore optimal policy insights, and serves as a nice completion to this paper.\medskip

\begin{enumerate}
\item Do not trust your own opinion and back your judgment until the action of the market itself confirms your opinion.
\item Markets are never wrong -- opinions often are.
\item The real money made in speculating has been in commitments showing in profit right from the start.
\item As long as a stock is acting right, and the market is right, do not be in a hurry to take profits.
\item The money lost by speculation alone is small compared with the gigantic sums lost by so-called investors who have let their investments ride.
\item Never buy a stock because it has had a big decline from its previous high.
\item Never sell a stock because it seems high-priced.
\item Never average losses.
\item Big movements take time to develop.
\item It is not good to be too curious about all the reasons behind price movements.
\end{enumerate}

\appendix
\section{}

We also observe that a \emph{Price Process}, denoted by $p_t(x_t)$,
will also satisfy a Bellman optimality; for risk neutral traders with information set $x_t$ and
dividends, or rewards, $ r_{t+1} (X_{t+1} )$, we have
\begin{equation*} 
p_t ( x_t ) = \max\nolimits_a \mathbb{E}^p \left[   r_{t+1} (X_{t+1} ) + p_{t+1} ( X_{t+1} ) \right], 
\end{equation*}
where $ \mathbb{E}^p[\cdot]$ denotes expectation with respect to the probability $p(S_{t+1}| s_t,a,x_t)$.

\end{document}